\documentclass[twocolumn,english,aps,prl,twocolums,showpacs,floatfix]{revtex4}
\usepackage[T1]{fontenc}
\usepackage[latin9]{inputenc}
\usepackage{float}
\usepackage{graphicx}
\usepackage{esint}

\makeatletter
\@ifundefined{textcolor}{}
{%
 \definecolor{BLACK}{gray}{0}
 \definecolor{WHITE}{gray}{1}
 \definecolor{RED}{rgb}{1,0,0}
 \definecolor{GREEN}{rgb}{0,1,0}
 \definecolor{BLUE}{rgb}{0,0,1}
 \definecolor{CYAN}{cmyk}{1,0,0,0}
 \definecolor{MAGENTA}{cmyk}{0,1,0,0}
 \definecolor{YELLOW}{cmyk}{0,0,1,0}
 }

\makeatother

\usepackage{babel}

\begin{document}

\preprint{This line only printed with preprint option}

\title{Time and ensemble averaging in time series analysis}

\author{Miroslaw Latka}

\email{Miroslaw.Latka@pwr.wroc.pl}

\affiliation{Institute of Biomedical Engineering, Wroclaw University of Technology,
Wybrzeze Wyspianskiego 27, 50-370 Wroclaw, Poland}

\email{Miroslaw.Latka@pwr.wroc.pl}

\author{Massimiliano Ignaccolo}

\email{mi8@phy.duke.edu}

\affiliation{Physics Department, Duke University, Durham, NC 27709, USA}

\author{Wojciech Jernajczyk}

\email{jernajcz@ipin.edu.pl }

\affiliation{Department of Clinical Neurophysiology, Institute of Psychiatry and
Neurology, Sobieskiego 9, 02-957 Warszawa, Poland}

\author{Bruce J. West}

\email{Bruce.J.West@us.army.mil}

\affiliation{Information Science Directorate, Army Research Office, P.O. Box 12211,
Research Triangle, NC 27709-2211, USA}

\date{04/13/2010}

\begin{abstract}
In many applications expectation values are calculated by partitioning
a single experimental time series into an ensemble of data segments
of equal length. Such single trajectory ensemble (STE) is a counterpart
to\emph{ }a multiple trajectory ensemble (MTE\emph{) }used whenever
independent measurements or realizations of a stochastic process are
available. The equivalence of STE and MTE for stationary systems was
postulated by Wang and Uhlenbeck in their classic paper on Brownian
motion (Rev. Mod. Phys. \textbf{17}, 323 (1945)) but surprisingly
has not yet been proved. Using the stationary and ergodic paradigm
of statistical physics -- the Ornstein-Uhlenbeck (OU) Langevin equation,
we revisit Wang and Uhlenbeck's postulate. In particular, we find
that the variance of the solution of this equation is \emph{different}
for these two ensembles. While the variance calculated using the MTE
quantifies the spreading of \emph{independent} trajectories originating
from the same initial point, the variance for STE measures the spreading
of two \emph{correlated} random walkers. Thus, STE and MTE refer to
two completely different dynamical processes. Guided by this interpretation,
we introduce a novel algorithm of partitioning a single trajectory
into a phenomenological ensemble, which we name a threshold trajectory
ensemble (TTE), that for an ergodic system is equivalent to MTE. We
find that in the cohort of healthy volunteers, the ratio of STE and
TTE asymptotic variances of stage 4 sleep electroencephalogram is
equal to 1.96 $\pm$ 0.04 which is in agreement with the theoretically
predicted value of 2.
\end{abstract}
\maketitle
The ergodic hypothesis asserting the equivalence of time and ensemble
averages began with Boltzmann's \cite{Boltzmann1995} conjecture that
a single trajectory can densely cover a surface of constant energy
in phase space. His proof of the hypothesis as well as many subsequent
proofs were shown to be fatally flawed. It was not until metric decomposability
was introduced by Birkoff \cite{G.D.Birkoff1931} that a rigorous
mathematical theory of ergodicity began to take shape. Kinchin, who
wrote a seminal work on the mathematical foundations of statistical
mechanics \cite{Kinchin1949}, offered a surprisingly pragmatic approach
to the ergodic hypothesis. He proposed to sidestep potentially formidable
proofs of ergodicity and judge the theory constructed on such assumption
by its practical success or failure. This latter perspective is adopted
by the vast majority of physicists when averaging over independent
measurements is not possible.

Let us consider a stochastic process $X(t)$ that can be the time
series of the position of a particle undergoing the Brownian motion,
inter-beat interval of human heart or a plethora of other time series
generated by complex physical or physiological systems. In the absence
of direct evidence to the contrary, it is assumed that such processes
are ergodic so that their underlying dynamics can in principle be
deduced from a single, very often historical, record $X(t)$ measured
over a sufficiently long time. In the classic paper Wang and Uhlenbeck
wrote \cite{wang1945theory}:

...One can then cut the record in pieces of length \emph{T }(where
\emph{T }is long compared to all periods occurring in the process),
and one may consider the different pieces as the different records
of an ensemble of observations. In computing average values one has
in general to distinguish between an ensemble average and a time average.
However, for a stationary process these two ways of averaging will
always give the same result...

This quote describes a ubiquitous process of generating a \emph{phenomenological
ensemble} by partitioning a single dataset. In the context of time
series analysis, we also use the synonym\emph{ single trajectory ensemble
(STE)} as a counterpart to\emph{ }a\emph{ multiple trajectory ensemble
(MTE) }which is employed whenever independent measurements or realizations
of a stochastic process are available. Wang and Uhlenbeck have implicitly
linked the of validity of phenomenological (STE) ensembles to the
ergodicity of the underlying dynamical system.

\begin{figure}
\includegraphics[scale=0.9]{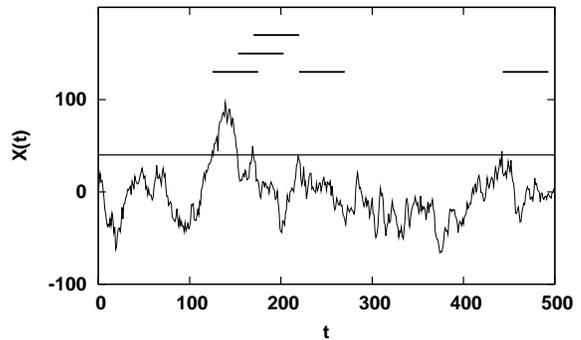}

\caption{An example of the solution $X(t)$ of the OU Langevin equation (\ref{langevin})
for $\lambda$$=$0.025 and $\sigma_{\eta}$$=$7.8. The horizontal
bars in the upper part of the figure represent those segments of the
displayed trajectory whose left endpoints are equal, with a predetermined
accuracy, to $X_{L}$. In other words, the left endpoints are the
intersection of the trajectory with the chosen threshold which is
marked in the graph by the horizontal gridline. Such segments of length
$\tau$ are used to construct a threshold trajectory ensemble discussed
later in the text.\label{fig:Trajectory}}

\end{figure}

It is remarkable that Wang and Uhlenbeck's postulate of equivalency
of STE and MTE, which in fact provided justification of a half a century
of empirical analyses, has not, to our knowledge, been directly tested.
Motivated by our recent study \cite{Massi2010PRE}, we revisit this
postulate using the stationary and ergodic paradigm of statistical
physics -- the Ornstein-Uhlenbeck (OU) Langevin equation:

\begin{equation}
\frac{dX(t)}{dt}=-\lambda X(t)+\eta\left(t\right)\label{langevin}\end{equation}
where $\lambda$ is the dissipation rate \cite{lindenberg}. In the
above equation, a zero-centered Gaussian random force $\eta\left(t\right)$
is delta correlated in time

\begin{equation}
\left\langle \eta\left(t\right)\eta\left(t+\tau\right)\right\rangle =\sigma_{\eta}^{2}\delta\left(\tau\right)\label{etaCorr}\end{equation}
and the angular brackets denote an average over an ensemble of realizations
of the random force. $X(t)$ is also a Gaussian random process with
a spectrum \cite{wang1945theory}: \begin{equation}
G(f)=\frac{2\sigma_{\eta}^{2}}{\lambda^{2}+4\pi^{2}f^{2}}.\label{LangevinSpectrum}\end{equation}
 Consequently, from the convolution theorem one obtains the autocorrelation
function of $X(t)$ \begin{equation}
\rho(t)=e^{-\lambda t}.\label{autocorrelation}\end{equation}
 $X(t)$ may be expressed as the formal solution to the first-order,
linear, stochastic differential equation (\ref{langevin}) \begin{equation}
X(t)=e^{-\lambda t}\left[X(0)+\int\limits _{0}^{t}\eta(t^{\prime})e^{\lambda t^{\prime}}dt^{\prime}\right].\label{langevinSolution}\end{equation}
In Fig. \ref{fig:Trajectory} we present a solution $X(t)$ generated
by numerical integration of Eq. (\ref{langevin}) with a constant
time step $\Delta t=1$ ($\lambda$$=$0.025 and $\sigma_{\eta}$$=$7.8).

\begin{figure}
\includegraphics[scale=0.9]{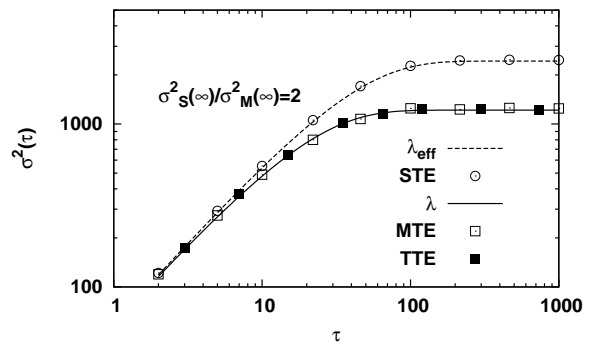}

\caption{Variance of the solution $X(t)$ of Eq. (\ref{langevin}) is plotted
as a function of the length of the data window $\tau$. The model's
parameters are the same as in Fig. \ref{fig:Trajectory}. The numerical
estimate of variance was calculated using the single (circles), multiple
(squares) and threshold (filled squares) trajectory ensembles. The
theoretical value of the variance is drawn for the MTE (solid line,
\emph{cf.} Eq. (\ref{langevinVariance})) and the STE (dashed line,
\emph{cf.} Eq. (\ref{varianceTrajectory2})). The sliding window algorithm
was used to generate the STE. The non-overlapping window partitioning
yielded the same results. The ratio of the asymptotic variances for
STE and MTE is is given by Eq. (\ref{eq:ratioVar}).\label{fig:STD}}

\end{figure}

According to Wang and Uhlenbeck's prescription, a long time series
generated by the OU Langevin equation can be partitioned to yield
a STE. We know that the solution to the OU Langevin equation is ergodic
so that averages obtained using STE and MTE should coincide. Variance
is the most commonly used measure of time series variability. Therefore,
let us perform computer simulations to calculate this metric for both
ensembles. In Fig. \ref{fig:STD}, $\sigma^{2}$ of the solution $X(t)$
of the Langevin equation (\ref{langevin}) is plotted as a function
of the length $\tau$ of the data window. The MTE variance $\sigma_{M}^{2}$
is denoted by opened squares and was calculated using $n_{M}=3000$
trajectories of length $N_{M}=1000$ originating at zero. The STE
variance $\sigma_{S}^{2}$, represented by open circles, was computed
by partitioning a single trajectory of length $N_{S}=n_{M}N_{M}$
into segments of length $\tau$ (we used a sliding window algorithm).
It is obvious that the two ways of calculating the variance are not
equivalent.

To clarify the difference between the two phenomenological ensembles
depicted in Fig. \ref{fig:STD}, we derive the analytical expressions
for the variance of the solution $X(t)$ for both ensembles. Let us
consider first the MTE of infinite number of trajectories all starting
from zero. Using Eq. (\ref{langevinSolution}) we obtain

\begin{equation}
\sigma_{M}^{2}\left(t\right)\equiv\left\langle X^{2}\left(t\right)\right\rangle _{M}-\left\langle X\left(t\right)\right\rangle _{M}^{2}=\frac{\sigma_{\eta}^{2}}{2\lambda}\left[1-e^{-2\lambda t}\right].\label{langevinVariance}\end{equation}
Please note that STE averaging, by its very nature, \emph{involves
relative displacements: $Z(t,\tau)=X(t+\tau)-X(t)$.} Using Eq\emph{.}~(\ref{langevinSolution}),
we can write $X(t+\tau)$ as \begin{equation}
X(t+\tau)=e^{-\lambda\tau}\left[X(t)+\int\limits _{0}^{\tau}\eta(t+t^{\prime})e^{\lambda t^{\prime}}dt^{\prime}\right]\label{shiftedLangevinSolution}\end{equation}
 and express $Z(t,\tau)$ in the following form \begin{equation}
Z(t,\tau)=e^{-\lambda\tau}\int\limits _{0}^{\tau}\eta(t+t^{\prime})e^{\lambda t^{\prime}}dt^{\prime}+X(t)\left[e^{-\lambda\tau}-1\right].\label{pseudoGibbs2}\end{equation}
 Recall that the random force $\eta$ is stationary and, therefore,
the time translation of $\eta$ in the integrand in Eq. (\ref{pseudoGibbs2})
does not affect the statistical properties of $Z(t,\tau)$. Without
loss of generality, we may assume that a trajectory $X(t)$ starts
at zero and then $Z(t,\tau)$ may be written in a particularly illuminating
form \begin{equation}
Z(t,\tau)=X(\tau)+X(t)\left[e^{-\lambda\tau}-1\right].\label{pseudoGibss4}\end{equation}
For a fixed value of $t$ (fixed left endpoint of the interval), the
relative displacement $Z(t,\tau)$ is a function of segment length
$\tau$. In particular, the first term on the r.h.s. of Eq. (\ref{pseudoGibss4})
is stochastic while the second one is purely deterministic. Consequently,
the covariance of these two terms vanishes. Thus, the partitioning
of a single trajectory is equivalent to building up \emph{deterministic
trends}. $Z(t,\tau)$ being the sum of Gaussian variables is a Gaussian
variable itself with zero mean value ($E[Z(t,\tau)]=0$) and the following
variance \begin{equation}
E[Z^{2}(t,\tau)]=\sigma_{M}^{2}(\tau)+\left[e^{-\lambda\tau}-1\right]^{2}\sigma_{M}^{2}(t).\label{Zvariance}\end{equation}
 The STE variance $\sigma_{S}^{2}(\tau)$ is just $E[Z^{2}(t,\tau)]$
time averaged along the trajectory of length $T$

\begin{eqnarray}
\sigma_{S}^{2}(\tau) & = & \frac{1}{T-\tau}\int\limits _{0}^{T-\tau}E[Z^{2}(t,\tau)]dt\label{varianceTrajectory}\\
 & = & \sigma_{M}^{2}(\tau)+\frac{(e^{-\lambda\tau}-1)^{2}\sigma_{\eta}^{2}}{2\lambda}\left[1+\frac{e^{-2\lambda(T-\tau)}-1}{2\lambda(T-\tau)}\right].\nonumber \end{eqnarray}
 Taking into account that $\lambda$$T$$\gg$1, we obtain the following
approximation

\begin{eqnarray}
\sigma_{S}^{2}(\tau) & \approx & \frac{\sigma_{\eta}^{2}}{\lambda}(1-e^{-\lambda\tau}).\label{varianceTrajectory2}\end{eqnarray}
Thus, the variance for the single trajectory ensemble is given by
the formula Eq. (\ref{langevinVariance}) for the MTE, albeit with
the effective dissipation rate $\lambda_{eff}=\lambda/2$ which is
half that of the MTE. Consequently, the ratio of the asymptotic variances
for STE and MTE is 

\begin{equation}
\sigma_{S}^{2}(\infty)/\sigma_{M}^{2}(\infty)=2.\label{eq:ratioVar}\end{equation}
Both curves $\sigma_{M}^{2}(\tau)$ and $\sigma_{S}^{2}(\tau)$ are
plotted in Fig. \ref{fig:STD} and are in agreement with the relevant
numerical calculations.

In hindsight, the observed disagreement between the two ways of calculating
the variance is less surprising than it ought to have been. $\sigma_{M}$
is the measure of spreading of statistically independent trajectories
that start at $X(0)$. On the other hand, the endpoints of intervals
used to calculate the relative displacements $Z(t,\tau)$ may be interpreted
as the final positions of two \emph{correlated} random walkers who
both start at $X(0)$ and whose correlation function $\rho(t)$ has
the exponential time dependence given by Eq. (\ref{autocorrelation}).
$\sigma_{S}(\tau)$ quantifies the spread of the distance between
such walkers after time $\tau$ and consequently refers to a completely
different dynamical process. The formal proof of this interpretation
is given elsewhere \cite{epl2010}.

Variance is certainly the most prevalent measure of time series variability.
Moreover, it is often a critical part of fractal scaling detection
algorithms, such as detrended fluctuation analysis (DFA) \cite{Peng1994,bashan2008comparison}.
In light of the difference between the STE and MTE variances, one
can easily envision the situation when simultaneous application of
both ensembles appears rational, but ultimately leads to systematic
errors. For example, one may perform the measurements on a cohort
of subjects to determine the variability of a physiological quantity.
However, when the variability determined for a given patient is compared
with that of the cohort, one may be inclined to improve the statistics
by averaging over a single trajectory ensemble. We know that for the
OU Langevin model, this approach leads to the gross overestimation
of the asymptotic variance. Thus, the question arises as to whether
it is possible to partition a single trajectory in such a way that
the resulting ensemble is equivalent to MTE and the improvement in
statistics is accomplished. 

The solution to this problem presents itself as soon as we realize
that in STE both endpoints of intervals contribute to spreading, whereas
in MTE all left endpoints  are the same initial condition. It is this
difference between the two ensembles that explains why the asymptotic
variance for STE is exactly twice that for MTE. The horizontal bars
in the upper part of Fig. \ref{fig:Trajectory} represent those segments
of the displayed trajectory whose left endpoints are equal, with a
predetermined accuracy $\epsilon$, to the $X_{L}$. In other words,
the left endpoints are the intersections of the trajectory with the
chosen threshold which is marked in the graph by the horizontal gridline.
Such segments of length $\tau$ are used to construct \emph{a threshold
trajectory ensemble (TTE). }The filled squares in Fig. \ref{fig:STD}
correspond to the variance $\sigma_{T}^{2}$ for such an ensemble.
The data segments with $X_{L}=0$ ($\epsilon=0.87$ which corresponds
to 2.5\% of the standard deviation of the trajectory of length $N_{S}=n_{M}N_{M}$
shown in Fig. \ref{fig:Trajectory}) were selected. There were approximately
60000 segments that satisfied the imposed criteria and, despite the
relatively small size of the TTE, the agreement with the MTE is apparent. 

\begin{figure}[H]
\includegraphics[scale=0.9]{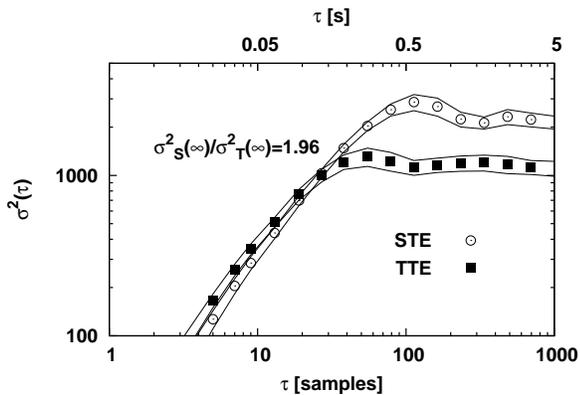}

\caption{The group averaged variance of stage 4 sleep electroencephalogram
is plotted as a function of data window length $\tau$ for STE (open
circles) and TTE (filled squares). The thin solid lines represent
the standard error of the mean. The sampling frequency of EEG was
200 Hz. The ratio of the asymptotic variances for STE and TTE 1.96
$\pm$ 0.04 is in agreement with the theoretical value given by Eq.
(\ref{eq:ratioVar}). \label{fig:varianceEEG}}

\end{figure}

In Fig. \ref{fig:varianceEEG} we present the group averaged variance
of stage 4 sleep electroencephalogram (EEG) plotted as a function
of data window length $\tau$ for STE (open circles) and TTE (filled
squares). The retrospective analysis was performed on data from 10
healthy volunteers. For each subject, we extracted a single 5 minute
data segment from the C3 channel of the polysomnogram \cite{niedermeyer04}.
$\epsilon$ was chosen as 2.5\% of the standard deviation $\sigma_{EEG}$
of EEG time series. For each volunteer, we averaged a variance curve
$\sigma_{T}^{2}(\tau)$ over the TTEs corresponding to the thresholds
spaced by $2\epsilon$ between $-2\sigma_{EEG}$ and $2\sigma_{EEG}$.
The single trajectory ensembles were constructed using the sliding
window partitioning to account for the fact that the intervals making
TTEs may overlap (\emph{cf.} Fig. \ref{fig:Trajectory}). The comparison
of Figs. \ref{fig:STD} and \ref{fig:varianceEEG} shows that both
for the model and for experimental data the initial growth of variance
is arrested. The OU Langevin models's parameters were determined via
a nonlinear fit of the group-averaged $\sigma_{T}^{2}$ to Eq. (\ref{langevinVariance}).
The ratio of STE and TTE asymptotic variances for EEG is equal to
1.96 $\pm$ 0.04 which is in agreement with the theoretically predicted
value of 2 given by Eq. (\ref{eq:ratioVar}). The asymptotic values
for the experimental data were obtained by averaging variance for
all intervals of length $\tau>1.5s$. As we already mentioned, the
value of theoretical ratio naturally stems from the two-walker interpretation
of STE. The application of the OU Langevin equation to EEG modeling
is described in detail in \cite{Massi2010PRE}.

The equivalency of MTE and STE averages was postulated as early as
in 1945 by Wang and Uhlenbeck \cite{wang1945theory}. Herein we demonstrated
the failure of this assumption for variance -- the most frequently
used measure of time series variability. By identifying the origin
of the failure, we were able to put forward a novel algorithm of partitioning
a single trajectory into a threshold trajectory ensemble that for
an ergodic system is equivalent to MTE. TTE obviates systematic errors
which may be introduced into data analysis by simultaneous application
of single and multiple trajectory ensembles.

Let us finish with a cautionary note. The ergodic hypothesis dates
back to the very beginning of statistical physics. The recent studies
\cite{Masuda2003,Brokmann2003,Margolin2005,Tang2005,Margolin2006,mccauley2008time,PhysRevLett.100.250602}
have once again brought ergodicity from the backstage into the limelight.
The growing list of non-ergodic systems should warn against indiscriminate
application of single trajectory (single particle) ensembles.

\bibliographystyle{apsrev}
\bibliography{ensembles}

\end{document}